\documentclass[twocolumn]{pp}

\usepackage[latin1]{inputenc}
\usepackage{amsmath}
\usepackage{amssymb}
\usepackage{amstext}
\usepackage{mathptmx}
\usepackage{amsfonts}
\usepackage{amsxtra}
\usepackage{amsbsy}
\usepackage{graphicx}

\newcommand{\N}{\mathbb N}

\newcommand{\R}{\mathbb R}

\newcommand{\de}{\partial}
\newcommand{\imp}{\operatorname{\mathrm{Im}}}

\newcommand{\leb}{\mathcal L}
\newcommand{\vc}[1]{\vec{#1}}
\newcommand{\vcd}[1]{\tilde{\vec{#1}}}
\newcommand{\dvg}{\operatorname{\mathrm{div}}}
\newcommand{\lap}{\Delta}
\newcommand{\ft}[1]{\widehat{#1}}
\newcommand{\eps}{\varepsilon}
\newcommand{\vt}[1]{\tens{#1}}
\newcommand{\J}{\mathcal J}

\newcommand{\I}{\mathcal I}
\newcommand{\Y}{\mathcal Y}
\newcommand{\K}{\mathcal K}

\newcommand{\tsp}{\dag}
\newcommand{\slet}{\vc{z}}

\newcommand{\sletp}{p_{\vc{z}}}
\newcommand{\tslet}{\vt Z}
\newcommand{\tsletc}{Z}
\newcommand{\ei}{\mathrm{Ei}}

\newcommand{\rod}{\Lambda}
\providecommand\bcdot{\boldsymbol{\cdot}}
\providecommand\bnabla{\boldsymbol{\nabla}}

\begin{document}

\title{Slender-body theory for viscous flow via dimensional reduction and hyperviscous regularization}

\author{Giulio G. Giusteri \and Eliot Fried}%

\institute{G. G. Giusteri \at
              Dipartimento di Matematica e Fisica ``N. Tartaglia'', Universit\`a Cattolica del Sacro Cuore, Brescia, 25121, Italy \\
              Tel.: +39-030-2406705\\
              Fax: +39-030-2406742\\
              \email{giulio.giusteri@unicatt.it}           
           \and
           E. Fried \at
              Mathematical Soft Matter Unit, Okinawa Institute of Technology Graduate University, 1919-1 Tancha, Onna-son, Kunigami, Okinawa 904-0495 Japan.
}

\date{\today}

\maketitle

\begin{abstract}
A new slender-body theory for viscous flow, based on the concepts of dimensional reduction and hyperviscous regularization, is presented. The geometry of flat, elongated, or point-like rigid bodies immersed in a viscous fluid is approximated by lower-dimensional objects, and a hyperviscous term is added to the flow equation. The hyperviscosity is given by the product of the ordinary viscosity with the square of a length that is shown to play the role of effective thickness of any lower-dimensional object. 
Explicit solutions of simple problems illustrate how the proposed method is able to represent with good approximation both the velocity field and the drag forces generated by rigid motions of the immersed bodies, in analogy with classical slender-body theories. This approach has the potential to open up the way to more effective computational techniques, since the complexity of the geometry can be significantly reduced. This, however, is achieved at the expense of involving higher-order derivatives of the velocity field. Importantly, both the dimensional reduction and the hyperviscous regularization, combined with suitable numerical schemes, can be used also in situations where inertia is not negligible.
\end{abstract}

\section{Introduction}

Composite systems where microscopic bodies move in a viscous fluid are ubiquitous in both biological and technological contexts, and the flows involved are often characterized by very small Reynolds numbers. In an early seminal contribution, Stokes \cite{Sto51} computed the flow past a translating rigid sphere in the low-Reynolds-number limit, determining the corresponding drag force and showing how Newtonian liquids can evade D'Alembert's paradox. This result was extended to ellipsoidal particle shapes by Oberbeck \cite{Obe76}, and to the case of rotating and shearing surrounding flows by Edwardes \cite{Edw92} and Jeffery \cite{Jef22}, respectively. The solution for a general surrounding flow was provided by Kim and Karrila \cite{KimKar05}.

An explicit solution of the same problem for a rigid body of general shape immersed in a low-Reynolds-number flow appears to be out of reach, but the broad spectrum of possible applications have favoured the development of methods to approximate the velocity field generated by the body, or at least the drag force and torque exerted on it. For a rigid body whose length is large compared with its breadth, starting with an idea of Burgers \cite{Bur38}, various authors, including Tuck \cite{Tuc64}, Tillett \cite{Til70}, Batchelor \cite{Bat70}, and Cox \cite{Cox70}, have contributed to the development of what is known as slender-body theory. In essence, that theory provides approximate expressions for the relevant quantities simulating the action of the immersed body on the surrounding fluid by means of force and torque distributions concentrated on suitable lines. The flow generated by those distributions is divergent precisely on the lines where forces concentrate, hence the use of the name ``singularity method'' to describe this technique.
The same methodology has been exploited also for particles which are not slender since the formative contribution of Oseen \cite{Ose27} and a clear account of it with applications to Stokes flows is given by Chwang \& Wu \cite{ChwWu75}.

Despite the possibility of approximating with good accuracy many quantities of physical interest, it is difficult to apply classical slender-body theory to particles of arbitrary shape and to situations featuring the presence of many particles, which are clearly important for applications. For this reason, modifications of the theory have been proposed by de~Mestre \cite{DeM73}, Johnson \cite{Joh80}, Barta and Liron \cite{BarLir88}, and Barta \cite{Bar11}, and also somewhat different theories have been developed by Lighthill \cite{Lig75,Lig76} and Keller and Rubinow \cite{KelRub76}.
In a recent paper, Cortez and Nicholas \cite{CorNic12} presented a new theory in which the concentrated force densities are replaced by forces which are localized in a small three-dimensional region lying within the slender body.
Our approach can be viewed as complementary to theirs, in that we consider not only concentrated force distributions, but also concentrated bodies, regularizing in turn the flow equation.

We propose a new slender-body theory which can be applied to rigid bodies whose slenderness can affect one dimension (flat bodies), two dimensions (elongated bodies), or even three dimensions (point-like spherical particles). In so doing, we are motivated by the understanding that the hyperviscous regularization of the Stokes equation described in \S\ref{sec:hypvis} makes it possible to obtain a solution for the flow past a translating point-like spherical particle such that: (i) the drag exerted by the particle on the fluid is equal to the total viscous traction exerted on a Newtonian fluid by a spherical particle; (ii) the solution of the regularized problem is a good approximation of the classical solution for a point-like sphere. Moreover, by comparing the classical and regularized solutions, it is possible to assign to the coefficient of the hyperviscous term the geometric meaning of effective thickness of the slender body, as discussed in \S\ref{sec:drhyp}.

In \S\ref{sec:greenf}, we derive the Green's function for the regularized Stokes operator, which is exploited, in \S\ref{sec:sph0}, to compute the flow generated by a point-like spherical particle, and to evaluate the hydrodynamic interaction between point-like spheres. The problem of the flow generated by a rigid rod is considered in \S\ref{sec:rod}, where some limitations of the singularity method, associated with the rigidity constraint, are considered. In the final discussion, we mention possible advantages in numerical simulations based on our theory, and we outline further developments.

\section{Hyperviscous regularization}\label{sec:hypvis}

The classical Stokes equation for incompressible Newtonian fluids, that is the low-Reynolds-number linearisation of the Navier--Stokes equation, reads
\begin{equation*}
\rho\frac{\de\vc u}{\de t}=-\bnabla p+\mu\lap\vc u+\rho\vc b\,,
\end{equation*}
where $p$ is the pressure field, $\vc u$ is the divergence-free velocity field, $\rho>0$ is the constant and homogeneous mass density, $\mu>0$ is the dynamic viscosity, and $\rho\vc b$ is a volumetric force density. For a steady flow, the Stokes equation reduces to
\begin{equation*}
\bnabla p-\mu\lap\vc u=\rho\vc b\,.
\end{equation*}

It is well-known that the regularity of solutions to the three-dimensional Navier--Stokes equation is still an open problem, and various modifications of that equation with better regularity theories have been analysed. Among those, the hyperviscous regularization (see Lions \cite{Lio69}, Chap.~I, Remarque 6.11) entails adding a term proportional to $\lap\lap\vc u$ to the equation. For this modified equation, the existence and uniqueness of regular solutions (that is of solutions which are continuous on the flow domain for every instant in a finite time interval) have been established. The low-Reynolds-number linearisation of that equation is, for a steady flow,
\begin{equation*}
\bnabla p-\mu\lap\vc u+\xi\lap\lap\vc u=\rho\vc b\,,
\end{equation*}
where the additional parameter $\xi>0$ is called the hyperviscosity.

Despite the mathematical appeal of the hyperviscous regularization, assigning a relevant physical meaning to $\xi$ can be problematic, other than that of a higher-order dissipation coefficient. In a series of papers, Fried and Gurtin \cite{FriGur06}, Musesti \cite{Mus09}, Giusteri \emph{et al.\ }\cite{GiuMar11}, and Giusteri \cite{Giu12} introduce and analyse different contributions to $\xi$ associated with dissipation functionals.
Here, we assign to $\xi$ a geometric, rather than dynamical, meaning, thereby allowing for a clearer interpretation in terms of standard physical quantities. 
To this end, we introduce a length-scale $L>0$, and set $\xi=\mu L^2$, so that the hyperviscous steady flow equation becomes 
\begin{equation}\label{eq:hypv}
\bnabla p-\mu\lap(\vc u-L^2\lap\vc u)=\rho\vc b\,.
\end{equation}

\subsection{Dimensional reduction and hyperviscosity}\label{sec:drhyp}

In the slender-body theory presented in this paper, the length scale $L$ is viewed as the effective thickness of the lower-dimensional objects which approximate three-dimensional bodies that are in one way or another slender. We now clarify how the hyperviscous regularization discussed above can be combined with dimensional reduction to approximate the flow past slender bodies, and elucidate the role of $L$ in this procedure.

Recall the Stokes problem of determining the steady flow $\vc v$ past a rigid sphere $\Sigma$ of radius $L$ centred at the origin,
\begin{equation}\label{pb:1}
\left\{
\begin{aligned}
\bnabla p_{\vc v}-\mu\lap\vc v=\vc 0\qquad&\text{in $\R^3\setminus\Sigma$,}\\
\dvg\vc v= 0\qquad&\text{in $\R^3\setminus\Sigma$,}\\
\vc v=\vc 0\qquad&\text{on $\de\Sigma$,}\\
\vc v=\vc U\qquad&\text{at infinity,}
\end{aligned}
\right.
\end{equation}
where $\vc U\in\R^3$ denotes the constant far-field velocity. The hyperviscous Stokes problem for the steady flow $\vc u$ past a point-like spherical particle placed at the origin has the form
\begin{equation}\label{pb:2}
\left\{
\begin{aligned}
\bnabla p_{\vc u}-\mu\lap(\vc u-L^2\lap\vc u)=\vc d\delta(\vc x)\qquad&\text{in $\R^3$,}\\
\dvg\vc u= 0\qquad&\text{in $\R^3$,}\\
\vc u=\vc 0\qquad&\text{at the origin,}\\
\end{aligned}
\right.
\end{equation}
where $\delta$ is the three-dimensional Dirac distribution and $\vc d$ is the drag force. Notice that a far-field condition akin to that entering~\eqref{pb:1} is not imposed in~\eqref{pb:2}. However,~\eqref{pb:2} admits a unique solution, and the value of $\vc u$ at infinity is determined by the force $\vc d$.

The solution to the Stokes equation~\eqref{pb:1} in the entirety of $\R^3$ is divergent at the origin; the solution of~\eqref{pb:2} (with $L>0$) is bounded and continuous in $\vc 0$, making explicit the reason why the hyperviscous equation~\eqref{eq:hypv} is considered a regularization of the classical one, namely~\eqref{eq:hypv} with $L=0$. It is important to notice that, in the limit $L\to 0$, the problem~\eqref{pb:2} becomes ill-posed.

In \S\ref{sec:sphere}, we solve explicitly problem~\eqref{pb:2}, showing that, far enough from the sphere of radius $L$, $\vc u$ is a good approximation of the solution $\vc v$ of problem~\eqref{pb:1}. Moreover, $\vc u$ takes at the origin the same value taken by $\vc v$ on the spherical surface $\de\Sigma$, and it turns out that, to obtain also $\vc u=\vc U$ at infinity, we must take the drag force exerted by the point-like spherical particle as
\[
\vc d=-\mu\int_{\de\Sigma}\left(\bnabla\vc v+\bnabla\vc v^\tsp\right)\vc n=6\pi\mu L\vc U\,,
\]
where $\vc n$ denotes the unit outward normal to $\de \Sigma$ and $\bnabla\vc v^\tsp$ denotes the transpose of $\bnabla\vc v$.
Namely, $\vc d$ must be equal to the traction exerted on the fluid by the sphere in problem~\eqref{pb:1}.
It is precisely this last fact which suggests the interpretation of $L$ as the effective thickness of the point-like spherical particle.

A point-like spherical particle represents a body which is slender in all spatial directions, but we are interested also in describing the motion of flat and elongated bodies. Actually, there are two different classes of problems worthy of mention: resistance problems and mobility problems.
\begin{itemize}
\item In a resistance problem, the velocities of the immersed objects are assigned and the goal is to determine the drag force required to sustain the motion.
\item In a mobility problem, the forces acting on the immersed objects are prescribed and the goal is to determine the resulting velocity field.
\end{itemize}

The solution of resistance and mobility problems in the case of a point-like spherical particle provides a linear relation between the assigned velocity and the drag force, which is the only force acting on the point-like spherical particle. The situation is remarkably different when dealing with flat or elongated rigid bodies. Whereas in resistance problems the prescription of the velocity of an immersed body automatically includes the rigidity constraint, the same constraint can be imposed in mobility problems only by assigning as force density the sum of the drag force and the reactive forces that are generated within an immersed body as a consequence of the assumption that it is rigid. We comment further on this in \S\ref{sec:rod}.

\section{Green's function for Stokes flow}\label{sec:greenf}

The basic tool used to construct solutions to the Stokes problem is the Stokeslet, that is the Green's function for the Stokes operator in $\R^3$. Its classical expression is given by
\begin{equation*}
p_{\vc s}(\vc x-\vc x')=\frac{\vc h\bcdot(\vc x-\vc x')}{4\pi|\vc x-\vc x'|^3}\,,
\end{equation*}
\begin{equation*}
\vc s(\vc x-\vc x')=\vt S(\vc x-\vc x')\vc h=\frac{\vc h}{8\pi\mu|\vc x-\vc x'|}+\frac{[\vc h\bcdot(\vc x-\vc x')](\vc x-\vc x')}{8\pi\mu|\vc x-\vc x'|^3}\,,
\end{equation*}
where $\vc h\in\R^3$ is any fixed vector and $\vt S$ is the Oseen tensor, with Cartesian components
\begin{equation*}
S_{ij}(\vc x-\vc x'):=\frac{\delta_{ij}}{8\pi\mu|\vc x-\vc x'|}+\frac{(x_i-x'_i)(x_j-x'_j)}{8\pi\mu|\vc x-\vc x'|^3}\,,
\end{equation*}
where $\delta_{ij}$ is Kronecker's symbol.

To compute the regularized Stokeslet, we first need a Green's function solution $g$ of the fourth-order elliptic equation
\[
\mu L^2\lap\lap g -\mu\lap g=\delta(\vc x-\vc x')\,.
\]
Using the Fourier transform, we easily obtain
\[
g(\vc x-\vc x')=\frac{1}{(2\pi)^3}\int_{\R^3} \frac{e^{i\vc k\bcdot(\vc x-\vc x')}}{\mu|\vc k|^2(L^2|\vc k|^2+1)}\,d\leb^3(\vc k)\,.
\]
We choose a basis for the momentum space in such a way that $\vc x-\vc x'$ is along the $k_3$-direction, set $R=|\vc x-\vc x'|$, switch to polar coordinates $(k,\theta,\phi)$, and use the calculus of residues to obtain
\begin{align*}
g(\vc x-\vc x')&\mbox{}=\frac{2\pi}{(2\pi)^3\mu L^2}\int_0^{+\infty}\int_{-1}^1\frac{e^{ikR\cos\theta}}{k^2+1/L^2}\,d(\cos\theta)\,dk\\
&\mbox{}=\frac{2}{(2\pi)^2\mu L^2 R}\int_0^{+\infty}\frac{\sin{kR}}{k(k^2+1/L^2)}\,dk\\
&\mbox{}=\frac{1}{(2\pi)^2\mu L^2 R}\imp\left[\int_{-\infty}^{+\infty}\frac{e^{ikR}}{k(k^2+1/L^2)}\,dk\right]\\
&\mbox{}=\frac{1}{(2\pi)^2\mu R}\left(\pi-\pi e^{-\frac{R}{L}}\right).
\end{align*}
Hence, the desired Green's function is
\begin{equation}\label{eq:g}
g(\vc x-\vc x')=\frac{1}{4\pi\mu|\vc x-\vc x'|}\left[1-\exp\left(-\frac{|\vc x-\vc x'|}{L}\right)\right]\,.
\end{equation}
Notice that, in the limit $L\to 0$, \eqref{eq:g} reduces to the Green's function
\begin{equation}\label{eq:g1}
g_1(\vc x-\vc x')=\frac{1}{4\pi\mu|\vc x-\vc x'|}
\end{equation}
for the Laplace equation.

\subsection{Regularized Stokeslet}\label{sec:slet}

We now proceed to construct the regularized Stokeslet, that is a pressure field $\sletp$ and a velocity field $\slet$ satisfying
\begin{equation}\label{eq:div0}
\left\{
\begin{aligned}
\dvg\slet &\mbox{}=0\,,\\
\bnabla\sletp+\mathcal A\slet &\mbox{}=\vc h\delta(\vc x-\vc x')\,,
\end{aligned}\right.
\end{equation}
with $\vc h\in\R^3$ and $\mathcal A=\mu L^2\lap\lap -\mu\lap$.

Let $\phi$ satisfy $\mathcal A\phi=\delta(\vc x-\vc x')$. Then, since $\mathcal A$ commutes with $\bnabla$, a solution for~\eqref{eq:div0} is given by
\[
\sletp=-\mathcal A\vartheta\,,
\]
\[
\slet=\vc h\,\phi+\bnabla\vartheta\,.
\]
The scalar field $\vartheta$ entering this solution is chosen to satisfy the divergence-free constraint and turns out to have the explicit form
\[
\vartheta=(-\lap)^{-1}(\vc h\bcdot\bnabla \phi)=g_1*(\vc h\bcdot\bnabla\phi)\,,
\]
where $g_1$ is as defined in \eqref{eq:g1}.

Now, exploiting the properties of the convolution and of the operator $\mathcal A$, and using the Green's function $g$ given by equation~\eqref{eq:g}, we find that
\begin{multline*}
-(g_1*\mathcal A(\vc h\bcdot\bnabla g))=-(g_1*\dvg(\mathcal A(g\vc h)))\\=-\dvg(g_1*\mathcal A(g\vc h))=-\vc h\bcdot\bnabla g_1\,
\end{multline*}
and, with a procedure similar to that leading to~\eqref{eq:g},
\begin{align*}
g_1*&(\vc h\bcdot\bnabla g)\mbox{}=\frac{1}{(2\pi)^3}\int i(\vc h\bcdot\vc k)\ft{g_1}\ft{g}e^{i\vc k\bcdot\vcd x}\,d\leb^3(\vc k)\\
&\mbox{}=\frac{1}{(2\pi)^3}\int\frac{i(\vc h\bcdot\vc k)e^{i\vc k\bcdot\vcd x}}{\mu|\vc k|^4(L^2|\vc k|^2+1)}\,d\leb^3(\vc k)\\
&\mbox{}
=\frac{\vc h\bcdot\vcd x}{4\pi^2\mu L^2|\vcd x|}\int_0^{+\infty}\int_{-1}^{1}\frac{i\cos\theta e^{ik|\vcd x|\cos\theta}}{k(k^2+1/L^2)}\,d(\cos\theta)\,dk\\
&\mbox{}=\frac{-\vc h\bcdot\vcd x}{4\pi^2\mu L^2|\vcd x|}\int_{-1}^{1}\tau\int_0^{+\infty}\frac{\sin(k|\vcd x|\tau)}{k(k^2+1/L^2)} \,dk \,d\tau\\
&\mbox{}
=\frac{-\vc h\bcdot\vcd x}{8\pi\mu|\vcd x|}\int_{-1}^{1}|\tau|\left(1-e^{-\frac{|\vcd x|}{L}|\tau|} \right)\,d\tau\\
&\mbox{}=-\frac{\vc h\bcdot\vcd x}{8\pi\mu|\vcd x|}\left[1+\frac{2L}{|\vcd x|}e^{-\frac{|\vcd x|}{L}}+\frac{2L^2}{|\vcd x|^2}\left(e^{-\frac{|\vcd x|}{L}}-1\right)\right]
\,,
\end{align*}
where we used the notation $\vcd x=\vc x-\vc x'$.

Defining now, for $r\in\R$, the functions
\[
\Sigma_1(r):=1-2e^{-r}-\frac{2}{r}e^{-r}-\frac{2}{r^2}\left(e^{-r}-1\right)
\]
and
\[
\Sigma_2(r):=1+2e^{-r}+\frac{6}{r}e^{-r}+\frac{6}{r^2}\left(e^{-r}-1\right)\,,
\]
we can write our regularized Stokeslet as
\begin{equation*}
\sletp(\vcd x)=\frac{\vc h\bcdot\vcd x}{4\pi|\vcd x|^3}\,,
\end{equation*}
\begin{equation*}
\slet(\vcd x)=\frac{\vc h}{8\pi\mu|\vcd x|}\Sigma_1(|\vcd x|/L)
+\frac{(\vc h\bcdot\vcd x)\vcd x}{8\pi\mu|\vcd x|^3}\Sigma_2(|\vcd x|/L)
\,.
\end{equation*}
Notice that the regularization does not alter the pressure field. Moreover, the velocity field $\slet$ is well-defined for any $\vcd x\in\R^3$, as is the Green's function $g$, at variance with the classical expression for $\vc s$, which is singular at the origin.

We also define the tensor field $\tslet$, the regularized Oseen tensor, such that, using Cartesian components,
\begin{equation}\label{eq:OT}
\tsletc_{ij}(\vcd x):=\frac{\delta_{ij}}{8\pi\mu|\vcd x|}\Sigma_1(|\vcd x|/L)
+\frac{\tilde x_i\tilde x_j}{8\pi\mu|\vcd x|^3}\Sigma_2(|\vcd x|/L)
\,,
\end{equation}
whereby it follows 
that $\slet(\vcd x)=\tslet(\vcd x)\vc h$.
The Stokeslet allows us to obtain an integral representation for the solution of~\eqref{eq:hypv}, with vanishing condition at infinity, in the form of a convolution:
\begin{equation}\label{eq:convol}
\vc u(\vc x):=\rho\int_{\R^3} \tslet(\vc x-\vc x')\vc b(\vc x')\,d\leb^3(\vc x')\,.
\end{equation}

Due to the regularity of the tensor $\tslet$ at the origin and the properties of convolution, $\vc u$ as determined by~\eqref{eq:convol} is bounded and continuous on all of space, even if the force density $\rho\vc b$ is localized on a set of zero Lebesgue measure---for example, if $\vc b$ is a measure concentrated on a lower-dimensional subset of $\R^3$, \eqref{eq:convol} determines a well-behaved velocity field.
This makes it possible to model interactions between the fluid and lower-dimensional objects by assigning force distribution concentrated on them, and thereby to turn the approximate singularity solutions of the classical theory into exact solutions of an approximate model.
Notice that although~\eqref{eq:convol} readily offers a solution for mobility problems, the force density $\vc b$ must represent the interactions taking place in all of space, not only in the fluid region; knowledge of both the drag force applied on the rigid body and the reactive forces generated within the body is therefore required.

\section{Flows generated by point-like spherical particles}\label{sec:sph0}

\subsection{A single point-like spherical particle}\label{sec:sphere}

The mobility problem for a point-like spherical particle subjected to a drag force density $\vc d\delta(\vc x)$ (viewed in a reference frame in which the point-like spherical particle is at rest) can be readily solved on applying formula~\eqref{eq:convol}, and doing so gives the velocity field
\begin{equation}\label{eq:A}
\vc u(\vc x)=\tslet(\vc x)\vc d-\tslet(\vc 0)\vc d=\tslet(\vc x)\vc d-(6\pi\mu L)^{-1}\vc d\,.
\end{equation}

The resistance problem requires a bit more of work. Recall that the disturbance field generated by a sphere rigidly translating with constant velocity $\vc U$ in a Newtonian liquid is given by (see \emph{e.g.\ }\cite{KimKar05})
\begin{equation}\label{eq:C}
\vc u^s(\vc x)=-a\frac{\vc U}{|\vc x|}-b\frac{\vc U}{|\vc x|^3}-a\frac{(\vc U\bcdot\vc x)\vc x}{|\vc x|^3}+b\frac{3(\vc U\bcdot\vc x)\vc x}{|\vc x|^5}\,,
\end{equation}
where the constants $a$ and $b$ can be determined by imposing the no-slip condition $\vc u^s=\vc U$ on the surface of any sphere of finite radius. However, since $\vc u^s$ is divergent at the origin, we cannot impose the no-slip condition on a point-like spherical particle, namely a sphere with vanishing radius.

For a hyperviscous liquid, we seek the divergence-free disturbance field $\vc u$ produced by a translating sphere. It satisfies the equation
\begin{equation}\label{eq:usphere}
\bnabla p-\mu\lap(\vc u-L^2\lap\vc u)=\vc 0\,.
\end{equation}
Since, given a velocity field $\vc u^s$ defined on all of space, it is always possible to find a pressure field $p^s$ such that $\bnabla p^s-\mu\lap\vc u^s=\vc 0$, any solution of
\begin{equation}\label{eq:helm}
\vc u-L^2\lap\vc u=\vc u^s
\end{equation}
provides a solution for~\eqref{eq:usphere}, with pressure $p^s$. It is readily checked that $\vc u^g=\vc u^s+L^2\lap\vc u^s$ is a particular solution of~\eqref{eq:helm}. We write the general solution of the associated homogeneous equation, using spherical coordinates $(r,\theta,\phi)$, as
\[
\vc u^o=\sum_{n=0}^{+\infty}\sum_{l=-n}^{n}\vc c^{n,l}\,h^{(1)}_{n}(ir/L)\,Y_n^l(\theta,\phi)\,,
\]
where $\vc c^{n,l}$ are vectors of coefficients, $h^{(1)}_{n}$ are spherical Hankel functions of the first kind, and $Y_n^l$ are spherical harmonics. Any solution of~\eqref{eq:usphere} is then of the form $\vc u=\vc u^g+\vc u^o$, and the boundary conditions determine the coefficients.

We now wish to show that it is possible to obtain a disturbance field that is bounded at the origin and satisfies the no-slip condition on a sphere with vanishing radius. We assume without loss of generality that $\vc U=U_3\vc e_3$, and we write the Cartesian components of $\vc u$ in spherical coordinates (the angular dependence of $Y_n^l$ being understood), obtaining:
\begin{multline*}
u_1(r,\theta,\phi)=U_3\left(\frac{3b+6aL^2}{r^3}-\frac{a}{r}\right)\sqrt{\frac{2\pi}{15}}\left(Y_2^{-1}-Y_2^1\right)\\+\sum_{n=0}^{+\infty}\sum_{l=-n}^{n}c_1^{n,l}\,h^{(1)}_{n}(ir/L)\,Y_n^l\,;
\end{multline*}
\begin{multline*}
u_2(r,\theta,\phi)=iU_3\left(\frac{3b+6aL^2}{r^3}-\frac{a}{r}\right)\sqrt{\frac{2\pi}{15}}\left(Y_2^{-1}+Y_2^1\right)\\+\sum_{n=0}^{+\infty}\sum_{l=-n}^{n}c_2^{n,l}\,h^{(1)}_{n}(ir/L)\,Y_n^l\,;
\end{multline*}
\begin{multline*}
u_3(r,\theta,\phi)=\frac{U_3}{3}\left(\frac{3b+6aL^2}{r^3}-\frac{a}{r}\right)\sqrt{\frac{16\pi}{5}}Y_2^0\\-\frac{U_3 2a\sqrt{16\pi}}{3r}Y_0^0+\sum_{n=0}^{+\infty}\sum_{l=-n}^{n}c_3^{n,l}\,h^{(1)}_{n}(ir/L)\,Y_n^l\,.
\end{multline*}
Thanks to the orthogonality of the spherical harmonics and the properties of the spherical Hankel functions, imposing $u_1(0,\theta,\phi)=u_2(0,\theta,\phi)=0$, we find, for any value of $a$, that $b=0$ and that the other non-vanishing coefficients are given by
\[
c_1^{2,1}=-c_1^{2,-1}=ic_2^{2,-1}=ic_2^{2,1}=\frac{2U_3a}{L}\sqrt{\frac{2\pi}{15}}\,.
\]
The third condition, namely $u_3(0,\theta,\phi)=U_3$, implies that $c_3^{0,0}=U_3\sqrt{4\pi}$, $c_3^{2,0}=U_3\sqrt{4\pi/5}$, and $a=-3L/4$, while all the remaining $c_3^{n,l}$ vanish.

Hence, the disturbance field is given by
\begin{equation*}
u_1(\vc x)=\frac{3L}{4}\frac{U_3x_3x_1}{|\vc x|^3}\Sigma_2(|\vc x|/L)
\,,
\end{equation*}
\begin{equation*}
u_2(\vc x)=\frac{3L}{4}\frac{U_3x_3x_2}{|\vc x|^3}\Sigma_2(|\vc x|/L)
\,,
\end{equation*}
\begin{equation*}
u_3(\vc x)=\frac{3L}{4}\frac{U_3}{|\vc x|}\Sigma_1(|\vc x|/L)
+\frac{3L}{4}\frac{U_3x_3^2}{|\vc x|^3}\Sigma_2(|\vc x|/L)
\,.
\end{equation*}
We thus find that the velocity field, in a reference frame in which the point-like spherical particle is at rest, is
\begin{equation}\label{eq:B}
\vc u(\vc x)=6\pi\mu L\tslet(\vc x)\vc U-\vc U\,.
\end{equation}
Comparing~\eqref{eq:A} and~\eqref{eq:B}, we conclude that the drag force needed to sustain the latter motion is given by the celebrated Stokes relation $\vc d=6\pi\mu L\vc U$.
Finally, when $|\vc x|> L$, the difference between the velocity field given by~\eqref{eq:B} and the classical solution of~\eqref{pb:1}, obtained by~\eqref{eq:C} with $a=-3L/4$ and $b=-L^3/4$, is small and decays as $|\vc x|^{-3}$ in the far field.


\subsection{Two point-like spherical particles}\label{sec:2part}

For the resistance problem related to the motion of two point-like spherical particles translating with constant velocity, it is tempting to exploit the correspondence developed in \S\ref{sec:sphere}, and solve instead the mobility problem, using~\eqref{eq:convol} and the superposition principle. That procedure works only when the two point-like spheres move with the same velocity, because the flow produced by the motion of the particles must be steady. 

Let us consider the disturbance field $\vc w$ produced by the two particles translating with velocity $\vc U$. We apply the drag force density
\[
\rho\vc b(\vc x)=\vc d_1\delta(\vc x-\vc y_1)+\vc d_2\delta(\vc x-\vc y_2)
\]
to the system, where $\vc y_1$ and $\vc y_2$ are the positions of the particles in a co-moving frame. Now the solution of the mobility problem reads
\begin{equation*}
\vc w(\vc x)=\tslet(\vc x-\vc y_1)\vc d_1+\tslet(\vc x-\vc y_2)\vc d_2\,,
\end{equation*}
and, to find the solution for the resistance problem, we must be able to choose $\vc d_1$ and $\vc d_2$ in such a way that $\vc w(\vc y_1)=\vc w(\vc y_2)=\vc U$. We have the conditions
\[
\tslet(\vc 0)\vc d_1+\tslet(\vc y_1-\vc y_2)\vc d_2=\vc U=\tslet(\vc 0)\vc d_2+\tslet(\vc y_1-\vc y_2)\vc d_1\,,
\]
which readily imply $\vc d_1=\vc d_2=:\vc d_0$, and
\begin{equation*}
\vc d_0=[\tslet(\vc 0)+\tslet(\vc y_1-\vc y_2)]^{-1}\vc U\,.
\end{equation*}
Without loss of generality, we may take one of the coordinate axes to be parallel to $\vc y_1-\vc y_2$, and we can easily see that $\vc d_0$ is parallel to $\vc U$ only when the latter is either parallel or orthogonal to $\vc y_1-\vc y_2$. In any other case, we find that a component of the force which is not in the direction of the motion is needed. That additional component of force is required to cancel the hydrodynamic interaction which tends to move the points away from the desired trajectory, since the leading point (the one in front with respect to the direction of $\vc U$) pulls and is pushed by the other point with a force directed along $\vc y_1-\vc y_2$.

\subsection{Hydrodynamic interaction between point-like spherical particles}\label{sec:npart}

Even though the above techniques cannot provide the full solution to the unsteady situation in which two point-like spherical particles move with different velocities, it is possible to utilize a procedure, known as method of reflections (see Kim and Karrila \cite{KimKar05}), to approximate the instantaneous flow field and the instantaneous hydrodynamic interaction between the particles.
So far, we have considered particles moving in a fluid which is at rest at infinity, but it is easy to include the effect of more complicated conditions in the far-field, as described by a field $\vc w_{\infty}$. Indeed, the disturbance field produced by a particle uniformly translating with velocity $\vc U$ in an ambient flow $\vc w_{\infty}$ is
\[
\vc u(\vc x)=6\pi\mu L\vt Z(\vc x-\vc y)[\vc U-\vc w_{\infty}(\vc y)]\,,
\]
where $\vc y$ is the position of the particle in a co-moving frame.

The method of reflections for two particles (labelled $1$ and $2$) consists of an iterative scheme which, at every step $k$, takes as ambient field for a particle $\vc w_{\infty}$ plus the field generated by the other particle in the step $k-1$.
Now we have $i,j=1,2$, $i\neq j$, and
\[
\vc u_i^{(0)}(\vc x)=6\pi\mu L\vt Z(\vc x-\vc y_i)[\vc U_i-\vc w_{\infty}(\vc y_i)]\,,
\]
\[
\vc u_i^{(k)}=\vt Z(\vc x-\vc y_i)\vc F^{(k)}_i\,,
\]
where $\vc U_1$ and $\vc U_2$ are the instantaneous velocities of the two particles, and
\[
\vc F^{(k)}_i=6\pi\mu L(\vc U_i-\vc w_{\infty}(\vc y_i)-\vt Z(\vc y_i-\vc y_j)\vc F^{(k-1)}_j)\,.
\]
Next, we wish to express $\vc F^{(2k)}_i$ and $\vc F^{(2k+1)}_i$ in terms of
\[
\vc F^{(0)}_1=6\pi\mu L(\vc U_1-\vc w_{\infty}(\vc y_1))
\]
and
\[
\vc F^{(0)}_2=6\pi\mu L(\vc U_2-\vc w_{\infty}(\vc y_2))\,.
\]
After some manipulations, we obtain
\begin{equation}\label{eq:int1}
\vc F^{(2k)}_i=\vt Q^{2k}\vc F^{(0)}_i +\sum_{n=0}^{k-1}\Big(	\vt Q^{2n}\vc F^{(0)}_i+\vt Q^{2n+1}\vc F^{(0)}_j\Big)\,
\end{equation}
and
\begin{equation}\label{eq:int2}
\vc F^{(2k+1)}_i=\vc F^{(0)}_i+\vt Q^{2k+1}\vc F^{(0)}_i +\sum_{n=0}^{k-1}\Big(\vt Q^{2n+2}\vc F^{(0)}_i+\vt Q^{2n+1}\vc F^{(0)}_j\Big)\,,
\end{equation}
where $\vt Q:=-6\pi\mu L\,\vt Z(\vc y_1-\vc y_2)$.

Notice that, given the expression~\eqref{eq:OT} of $\vt Z$, with every application of $\vt Q$ an additional factor of $L/|\vc y_1-\vc y_2|$ appears. It therefore becomes easy to keep track of the order of the approximation in terms of that small parameter: the $k$-th approximation $\vc F^{(k)}_i$ is exact up to the order $(L/|\vc y_1-\vc y_2|)^{k-1}$.
The same is not true if we extend the procedure to the case of $N$ point-like spherical particles, because of the appearance of corrections to lower-order terms even after many iterations. Indeed, let $N$ identical point-like spherical particles be placed at distinct points $\vc y_i$ moving with velocity $\vc U_i$, $i=1,\ldots,N$.
Taking $i,j=1,\ldots,N$, we have
\[
\vc u_i^{(0)}(\vc x)=6\pi\mu L\vt Z(\vc x-\vc y_i)[\vc U_i-\vc w_{\infty}(\vc y_i)]\,,
\]
\[
\vc F^{(0)}_i=6\pi\mu L(\vc U_i-\vc w_{\infty}(\vc y_i))\,,
\]
\[
\vc u_i^{(k)}=\vt Z(\vc x-\vc y_i)\vc F^{(k)}_i\,,
\]
where we have set
\[
\vc F^{(k)}_i=6\pi\mu L\Big(\vc U_i-\vc w_{\infty}(\vc y_i)-\sum_{j\neq i}\vt Z(\vc y_i-\vc y_j)\vc F^{(k-1)}_j\Big)\,.
\]
Considering now indexes $i_0,i_1,\ldots=1,\ldots,N$, we obtain
\begin{equation}\label{eq:int3}
\vc F^{(k)}_{i_0}=\vc F^{(0)}_{i_0}+\sum_{n=1}^{k}\left[\sum_{i_1,\ldots,i_n}\left(\prod_{m=1}^{n}\vt Q_{i_{m-1}i_m}\right)\vc F^{(0)}_{i_n}\right]\,,
\end{equation}
where $\vt Q_{kj}=\vt Q_{jk}:=-\delta_{jk}6\pi\mu L\,\vt Z(\vc y_j-\vc y_k)$, and the product in the above formula is the standard one arising in matrix multiplication.
Though somewhat complicated, expressions~\eqref{eq:int1}, \eqref{eq:int2}, and \eqref{eq:int3} can be used to approximate the hydrodynamic interactions in dilute suspensions of point-like spherical particles, that is suspensions in which the average separation between particles is much greater than $L$.

\section{Approximating a rigid rod}\label{sec:rod}

We now consider the case of a rigid straight rod, represented by a line segment of length $2a$, lying on the $x_1$-axis in the interval $[-a,a]$, when viewed in a co-moving frame.
We are interested in the steady disturbance field produced by a constant rigid motion of the rod, represented by the rigid velocity field
\[
\vc w(\vc x)=\vt\Omega\vc x+\vc U\,,
\]
where $\vt\Omega$ is the skew-symmetric spin (namely, a constant and uniform spin tensor), and where $\vc U$ is a constant translational velocity. 
We work in a co-moving frame and, as a consequence, the flow at infinity matches $-\vc w$. In this setting, we may assume that the steady Stokes flow has the form $\vc u-\vc w$, where $\vc u$ represents the disturbance flow, and is such that $\vc u=\vc w$ on the rod and $\vc u=\vc 0$ at infinity.

We are again facing a resistance problem; however, we first attempt to solve a mobility problem, introducing a force density concentrated on the rod $\rod:=[-a,a]\times\{(0,0)\}$ in the form
\[
\rho\vc b(\vc x)=\vc f(x_1)\delta_{\rod}(\vc x)=\vc f(x_1)\chi_{[-a,a]}(x_1)\delta(x_2)\delta(x_3)\,,
\]
where $\chi_{[-a,a]}(x_1)$ is the characteristic function of $[-a,a]$, and
$\delta_{\rod}$ represents a uniform probability measure supported on $\rod$. Using the convolution~\eqref{eq:convol}, we have
\begin{align*}
&u_i(\vc x)=\int_{\R^3}\sum_j\tsletc_{ij}(\vc x-\vc x')f_j(x'_1)\chi_{[-a,a]}(x'_1)\delta(x'_2)\delta(x'_3)\,d\vc x'\\
&=\int_{-a}^{a}\frac{f_i(x'_1)}{8\pi\mu M}\Sigma_1(M/L)\,dx'_1
\\
&\quad+\sum_j\int_{-a}^{a}\frac{(x_i-\delta_{1i}x'_i)(x_j-\delta_{1j}x'_j)f_j(x'_1)}{8\pi\mu M^3}\Sigma_2(M/L)\,dx'_1,
\end{align*}
where we have introduced $\sqrt{(x_1-x'_1)^2+x_2^2+x_3^2}=:M$.
We evaluate the previous expression at $x_2=x_3=0$, giving
\begin{align*}
u_i(x_1,0,0)&\mbox{}=\int_{x_1-a}^{x_1+a}\frac{f_i(x_1-y)}{8\pi\mu |y|}\Sigma_1(|y|/L)\,dy
\\
&\mbox{}+\delta_{1i}\int_{x_1-a}^{x_1+a}\frac{f_1(x_1-y)}{8\pi\mu |y|}\Sigma_2(|y|/L)\,dy\,.
\end{align*}
Since we are interested in the solution for $|x_1|\leq a$, to impose the matching with $w_i$ on $\rod$, we may split the integral into two parts, $u_i^+$ and $u_i^-$, accordingly evaluating $|y|$:
\begin{align*}
u_i^+(x_1)&\mbox{}=\int_{0}^{x_1+a}\frac{f_i(x_1-y)}{8\pi\mu y}\Sigma_1(y/L)\,dy
\\
&\mbox{}+\delta_{1i}\int_{0}^{x_1+a}\frac{f_1(x_1-y)}{8\pi\mu y}\Sigma_2(y/L)\,dy\,,
\end{align*}
\begin{align*}
u_i^-(x_1)&\mbox{}=\int_{x_1-a}^{0}\frac{-f_i(x_1-y)}{8\pi\mu y}\Sigma_1(-y/L)\,dy
\\
&\mbox{}+\delta_{1i}\int_{x_1-a}^{0}\frac{-f_1(x_1-y)}{8\pi\mu y}\Sigma_2(-y/L)\,dy\,.
\end{align*}
In agreement with classical slender-body theory, proposed, for instance, by Batchelor \cite{Bat70}, we make the following Ansatz on the concentrated force field:
\begin{equation}\label{eq:force}
f_i(\tau)=s_i\tau+c_i\,\qquad\text{with }s_i,c_i\in\R\,,\text{ for }i=1,2,3.
\end{equation}
In the sequel, we will need the following indefinite integrals, for $\alpha\in\R$ and $n\in\N$:
\[
\int\frac{1}{y^n}e^{\alpha y}\,dy=-e^{\alpha y}\sum_{k=1}^{n-1}\frac{(n-1-k)!}{(n-1)!}\frac{\alpha^{k-1}}{y^{n-k}}+\frac{\alpha^{n-1}}{(n-1)!}\ei(\alpha y)
\]
where $\ei$ denotes the exponential integral function, whose series expansion around $y=0$ has the form
\[
\ei(y)=\gamma+\log|y|+\sum_{k=1}^{+\infty}\frac{y^k}{k!\,k}\,,
\]
with $\gamma$ being the Euler--Mascheroni constant.

We next evaluate $u_i^+$ and $u_i^-$ for a force of the form~\eqref{eq:force} and find that they are given by
\begin{align*}
u_i^+(x_1)=\mbox{}&\frac{c_i+s_ix_1}{8\pi\mu}\left(\I_{0}^{x_1+a}+\delta_{1i}\hat{\I}_{0}^{x_1+a}\right)\\
&\mbox{}-\frac{s_i}{8\pi\mu}\left(\J_{0}^{x_1+a}+\delta_{1i}\hat{\J}_{0}^{x_1+a}\right)\,
\end{align*}
and
\begin{align*}
u_i^-(x_1)=\mbox{}&\frac{-(c_i+s_ix_1)}{8\pi\mu}\left(\Y^{0}_{x_1-a}+\delta_{1i}\hat{\Y}^{0}_{x_1-a}\right)\\
&\mbox{}+\frac{s_i}{8\pi\mu}\left(\K^{0}_{x_1-a}+\delta_{1i}\hat{\K}^{0}_{x_1-a}\right)\,,
\end{align*}
with
\[
\I_{0}^{x_1+a}=\left[\log y-\ei\Big[\frac{-y}{L}\Big]+\frac{L}{y}e^{-\frac{y}{L}}+\frac{L^2}{y^2}\left(e^{-\frac{y}{L}}-1\right)\right]_{0}^{x_1+a}\,,
\]
\[
\hat{\I}_{0}^{x_1+a}=\left[\log y-\ei\Big[\dfrac{-y}{L}\Big]-\frac{3L}{y}e^{-\frac{y}{L}}-\frac{3L^2}{y^2}\Big(e^{-\frac{y}{L}}-1\Big)\right]_{0}^{x_1+a},
\]
\[
\J_{0}^{x_1+a}=\left[y+2Le^{-\frac{y}{L}}+\frac{2L^2}{y}\left(e^{-\frac{y}{L}}-1\right)\right]_{0}^{x_1+a}\,,
\]
\[
\hat{\J}_{0}^{x_1+a}=\left[y-2Le^{-\frac{y}{L}}-\frac{6L^2}{y}\left(e^{-\frac{y}{L}}-1\right)\right]_{0}^{x_1+a}\,,
\]
\[
\Y^{0}_{x_1-a}=\left[\log |y|-\ei(y/L)-\frac{L}{y}e^{\frac{y}{L}}+\frac{L^2}{y^2}\left(e^{\frac{y}{L}}-1\right)\right]^{0}_{x_1-a}\,,
\]
\[
\hat{\Y}^{0}_{x_1-a}=\left[\log |y|-\ei(y/L)+\frac{3L}{y}e^{\frac{y}{L}}-\frac{3L^2}{y^2}\left(e^{\frac{y}{L}}-1\right)\right]^{0}_{x_1-a}\,,
\]
\[
\K^{0}_{x_1-a}=\left[y-2Le^{\frac{y}{L}}+\frac{2L^2}{y}\left(e^{\frac{y}{L}}-1\right)\right]^{0}_{x_1-a}\,,
\]
\[
\hat{\K}^{0}_{x_1-a}=\left[y+2Le^{\frac{y}{L}}-\frac{6L^2}{y}\left(e^{\frac{y}{L}}-1\right)\right]^{0}_{x_1-a}\,,
\]
where evaluation at $0$ stands for the limit, as necessary.
Now, we have
\begin{align}
u_i&(x_1,0,0)\mbox{}=u_i^+(x_1)+u_i^-(x_1)\notag\\
&\mbox{}=\frac{c_i+s_ix_1}{8\pi\mu}(\I_{0}^{x_1+a}+\delta_{1i}\hat{\I}_{0}^{x_1+a}-\Y^{0}_{x_1-a}-\delta_{1i}\hat{\Y}^{0}_{x_1-a})\notag\\
&-\frac{s_i}{8\pi\mu}(\J_{0}^{x_1+a}+\delta_{1i}\hat{\J}_{0}^{x_1+a}-\K^{0}_{x_1-a}-\delta_{1i}\hat{\K}^{0}_{x_1-a})\,.\label{eq:uwrong}
\end{align}

We next take $s_i=0$, as classical procedures would suggest, to analyse the case of rigid translations.
We compute the velocity in the middle of the rod, equate it to $\vc U$, and thereby arrive at the following approximation for the constant force densities, valid to any order in $\eps=L/2a$:
\begin{equation}
\left\{
\begin{aligned}
f_1(x'_1)&\mbox{}=c_1\approx\frac{2\pi\mu U_1}{\log(\eps^{-1})+\gamma-1/4+\eps^2}\,,\\
f_2(x'_1)&\mbox{}=c_2\approx\frac{4\pi\mu U_2}{\log(\eps^{-1})+\gamma+2-\eps^2}\,,\\
f_3(x'_1)&\mbox{}=c_3\approx\frac{4\pi\mu U_3}{\log(\eps^{-1})+\gamma+2-\eps^2}\,.
\end{aligned}\right.
\end{equation}
These expressions agree with results obtained by Batchelor \cite{Bat70}, but, unfortunately, the disturbance field given by~\eqref{eq:uwrong} is not constant on $\rod$, and hence $\vc u\neq\vc U$ except at the midpoint of the rod!

Indeed, such a solution is almost constant in a neighbourhood of the origin, and it would still be possible to impose a genuinely constant velocity on the boundary of a slender body of non-vanishing breadth, when $\rod$ lies well within it.
This justifies the classical approximation for slender bodies, but, since the object under consideration is one-dimensional, we actually need to obtain a velocity field which is exactly constant on the whole of $\rod$. As already mentioned, the previous discrepancy arises from the fact that we must include in the force density $\vc f$ the reactive force which guarantees the rigidity of the rod. An analogous result is obtained if we set $c_i=0$, to study a rigid rotation.

We might extract some insight regarding the form of such forces by actually solving the resistance problem for a rigidly translating rod, applying the procedure of \S\ref{sec:sphere}.
Indeed, with $\rod$ being the degenerate fundamental ellipse of a family of prolate spheroids,
we may view it also as a degenerate prolate spheroid.
The solution of this problem using prolate spheroidal coordinates and external spheroidal wave functions is provided in Appendix~\ref{app},
but that solution provides an expression for the disturbance field which is not readily expressed as a convolution.
To determine the reactive forces, we therefore rely on a different method.

For a fixed $N\in\N$, we approximate the rod by a sequence of equidistant points with first components $\{x_1^{k}\}_{k=1}^{2^N+1}$, chosen in $[-a,a]$ so that $x_1^{1}=-a$ and $x_1^{2^N+1}=a$. We then find the second component 
\[
\rho b_2(\vc x)=\delta(x_2)\delta(x_3)\sum_{k=1}^{2^N+1} f_2^{k}\delta(x_1^{k})\,
\]
of the force density required to make all the points move with equal velocity $\vc U=U_2\vc e_2$. (The same procedure produces analogous results for a velocity $\vc U=U_1\vc e_1$.)
The coefficients $f_2^k$, $k=1,2,\dots,2^N+1$, can be found by solving the Toeplitz system
\[
\begin{bmatrix}
a_0 & a_1 & a_2 & a_3 & \ddots & \ddots & a_{2^N} \\
a_1 & a_0 & a_1 & a_2 & \ddots & \ddots & \ddots \\
a_2 & a_1 & a_0 & a_1 & \ddots & \ddots & \ddots \\
a_3 & a_2 & a_1 & a_0 & \ddots & \ddots & \ddots \\
\ddots & \ddots & \ddots & \ddots & \ddots & \ddots&\ddots\\
\ddots & \ddots & \ddots & \ddots & \ddots & \ddots&a_1\\
a_{2^N} & \ddots & \ddots & \ddots & \ddots &a_1&a_0 
\end{bmatrix}
\begin{bmatrix}
\phantom{\vdots}f_2^1\\
\phantom{\vdots}f_2^2\\
\vdots\\
\vdots\\
\vdots\\
\vdots\\
\phantom{\vdots}f_2^{2^N+1}
\end{bmatrix}=
\begin{bmatrix}
\phantom{\vdots}U_2\\
\phantom{\vdots}U_2\\
\vdots\\
\vdots\\
\vdots\\
\vdots\\
\phantom{\vdots}U_2
\end{bmatrix}\,,
\]
with $a_i=\tsletc_{22}(ia/2^{N-1},0,0)$, for $i=0,1,\ldots,2^N$, which generalizes the technique used in \S\ref{sec:2part} for two point-like spherical particles.

\begin{figure*}
\begin{center}
\includegraphics[width=0.95\textwidth]{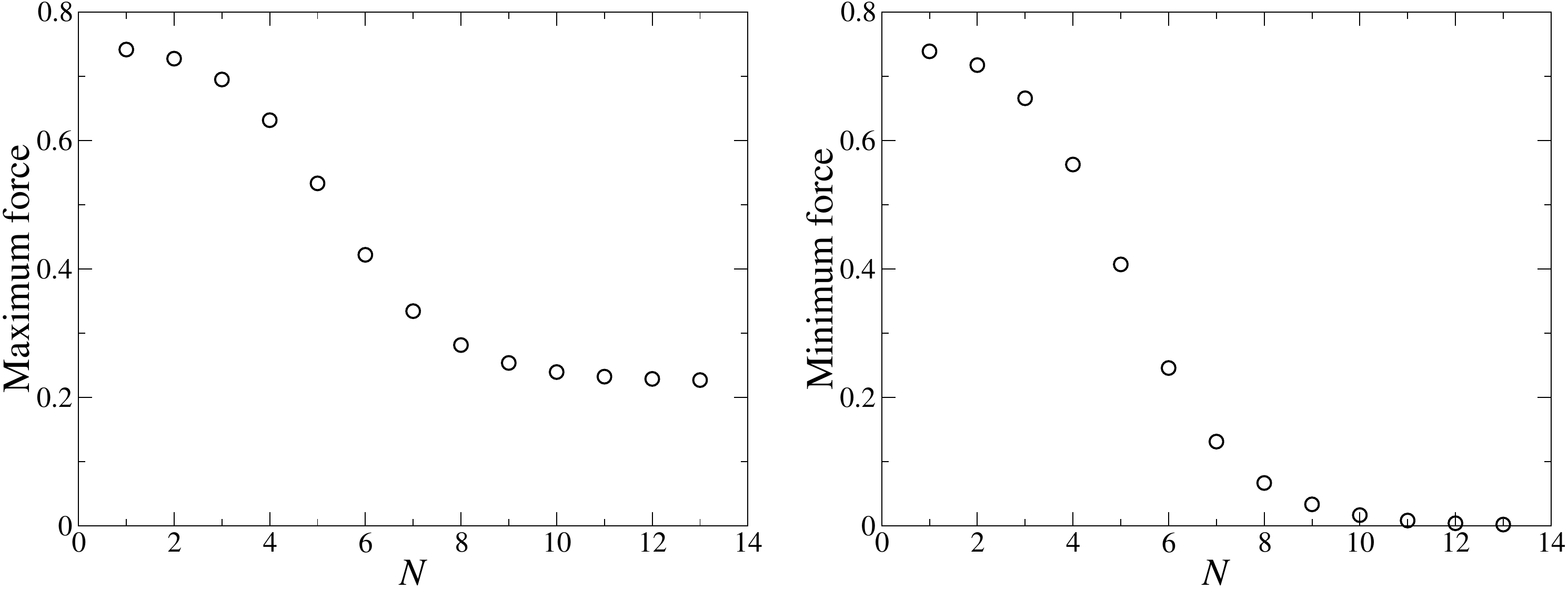}
\caption{Force coefficients $f_2^1$ (left) and $f_2^{2^{N-1}+1}$ (right) for different values of $N$.}\label{fig:minmax}
\end{center}
\end{figure*}

\begin{figure*}
\begin{center}
\includegraphics[width=0.95\textwidth]{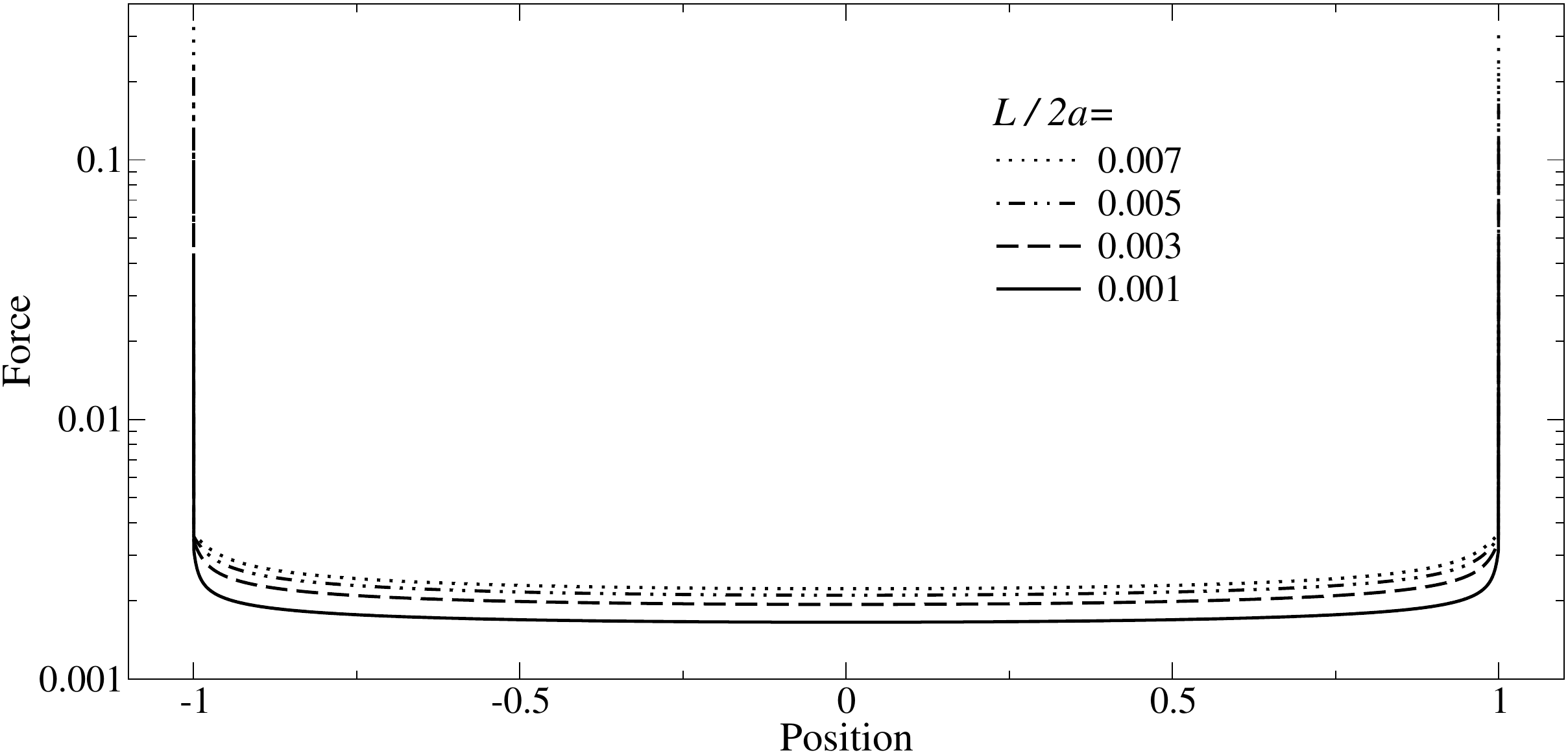}
\caption{Force density distributed on $\rod$ for different values of $L/2a$ (log-scale).}\label{fig:proL}
\end{center}
\end{figure*}

We solved the system for $N=1,2,\ldots,13$, fixing $a=1\:\mathrm{cm}$, $L/2a=0.005$, $U_2=1\:\mathrm{cm/s}$, and $\mu=4\:\mathrm{Pa\,s}$. Plots of the maximum and minimum coefficients, $f_2^1$ and $f_2^{2^{N-1}+1}$, against $N$ are provided in Figure~\ref{fig:minmax}, to show that the result approaches an asymptote for $N$ sufficiently large. The profile of the force density distributed on the rod, calculated for $a$, $U_2$, and $\mu$ as above and different values of $L/2a$ by using $8193$ points ($N=13$), is shown in Figure~\ref{fig:proL}. It is much larger (by two orders of magnitude) towards the ends of the rod, and rather flat in the middle, as expected, since the motion of the central points receives the maximum support from the hydrodynamic force generated by the motion of the other points, whereas the endpoints receive the least contribution from such interactions.
In accordance with the interpretation of $L$ as effective thickness, we observe that the magnitude of the force needed to sustain the motion increases with $L$.

\section{Summary and perspectives}\label{sec:conclusion}

We have presented a new slender-body theory, which, in contrast to previously developed alternatives, might be view\-ed as a modelling scheme, based on the concepts of dimensional reduction and hyperviscous regularization. We propose to approximate the geometry of flat, elongated, or point-like rigid bodies immersed in a viscous fluid with lower-dimensional objects, while adding a hyperviscous term to the flow equation. The hyperviscosity is given by the product of the ordinary viscosity with the square of a length that replaces the characteristic size of the body along the dimensions that are shrunken to zero in the slender-body limit.

We explicitly solved some simple problems to show that the proposed method is able to represent with good approximation both the velocity field and the drag forces generated by rigid motions of the immersed bodies, in analogy with classical slender-body theories. The regularized Stokeslet and the regularized Oseen tensor were determined. These objects are basic to the solution of mobility problems, in which the concentrated force density acting on the system is prescribed. We explained the limitations of this kind of solution technique, which stems from the need to determine the reactive force arising from the rigidity constraint. That force can be computed starting from the solution of resistance problems, in which the velocity of the immersed objects is prescribed.

The construction of a numerical scheme for the simulation of slender bodies in viscous fluids is a necessary step for any realistic application, and will be the object of a forthcoming paper. Within the present approach, the computational strategy stands to benefit from the understanding that the unsteady movement of the slender bodies is no longer represented by a time-varying domain, but rather by a time-dependent constraint on the function space of admissible velocity fields. The complexity of the geometry is therefore reduced at the expense of involving higher-order derivatives of the velocity field in the flow equation, and a time-dependent constraint on it. We nevertheless believe that this approach has the potential to enable more effective computational techniques, especially when dealing with suspensions featuring large numbers of particles.

Another important aspect of our approach is that both the dimensional reduction and the hyperviscous regularization, combined with suitable numerical schemes, can be used also in situations where inertia is not negligible. Indeed, as Giusteri \emph{et al.\ }\cite{GiuMar10} show, the flow generated by one-dimensional rigid bodies, moving in a fluid governed by a generalization of the Navier--Stokes equation with hyperviscosity, can be uniquely determined. A generalization to deformable bodies also seems feasible.\\

G.G.G.\ would like to express his gratitude to the Department of Mechanical Engineering of the University of Washington for the hospitality offered during the development of this research.

\appendix
\section{Uniform flow past a straight rod}\label{app}

To compute the low-Reynolds-number flow past a uniformly translating rigid rod, we follow the method of \S\ref{sec:sphere}. Specifically, we start from the solution $\vc u^{ps}$, given by Chwang and Wu \cite{ChwWu75}, for the disturbance field generated by a prolate spheroid with axis along $\vc e_1$ and foci at $x_1=\pm a$, uniformly translating with velocity $\vc U=U_1\vc e_1+U_2\vc e_2$. 
In view of the symmetry of the spheroid, there is no loss of generality in taking $U_3=0$. 
A particular solution of~\eqref{eq:helm} is $\vc u^{*}=\vc u^{ps}+L^2\lap\vc u^{ps}$. Setting
\[
r:=\sqrt{x_2^2+x_3^2}\,,\quad R_1:=\sqrt{(x_1+a)^2+r^2}\,,\quad R_2:=\sqrt{(x_1-a)^2+r^2}\,,
\]
\[
D_0:=\frac{R_2-(x_1-a)}{R_1-(x_1+a)}\,,\quad D_1:=\frac{1}{R_2}-\frac{1}{R_1}\,,\quad D_2:=\frac{x_1+a}{R_1}-\frac{x_1-a}{R_2}\,,
\]
\[
D_3:=\frac{1}{R_2^3}-\frac{1}{R_1^3}\,,\quad D_4:=\frac{x_1-R_2}{R_2^2-R_2(x_1-a)}-\frac{x_1-R_1}{R_1^2-R_1(x_1+a)}\,,
\]
\[
D_5:=\frac{1}{R_2^2-R_2(x_1-a)}-\frac{1}{R_1^2-R_1(x_1+a)}\,,
\]
we can express its components as
\begin{align*}
u^*_1(&\vc x)=\mbox{}-2(\alpha_1+\beta_1)\log D_0-2\beta_1 x_1D_1\\
&\mbox{}+2L^2\alpha_1 x_1D_3+(\beta_2x_2-2\beta_1x_1)D_4\\
&\mbox{}+\alpha_1D_2-\alpha_2 x_2D_1+\beta_2\frac{x_2}{r^2}(R_1-R_2)+2L^2\alpha_2\frac{x_2}{r^2}D_1\\
&\mbox{}+\frac{x_2x_1(x_1+a)}{r^2}\left(\frac{2L^2\alpha_2}{R_1^3}-\frac{\beta_2}{R_2}\right)+\frac{x_2x_1(x_1-a)}{r^2}\left(\frac{\beta_2}{R_1}-\frac{2L^2\alpha_2}{R_2^3}\right)\,,
\end{align*}
\begin{align*}
u^*_2(\vc x)=\mbox{}&(\beta_2-\alpha_2)\log D_0-(\alpha_1x_2+2\beta_1 x_2)D_1+2L^2\alpha_1 x_2D_3\\
&+\frac{x_1-a}{r^2}\left(\beta_2R_1+\frac{2L^2\alpha_2}{R_2}\right)
-\frac{x_1+a}{r^2}\left(\beta_2R_2+\frac{2L^2\alpha_2}{R_1}\right)\\
&+\frac{\alpha_2x_2^2D_2}{r^2}\left(1+\frac{4L^2}{r^2}\right)
+\frac{2L^2\alpha_2x_2^2}{r^2}\left(\frac{x_1+a}{R_1^3}-\frac{x_1-a}{R_2^3}\right)\\
&+\frac{\beta_2x_2^2}{r^2}\left[(x_1-a)\left(\frac{1}{R_1}-\frac{2R_1}{r^2}\right)
-(x_1+a)\left(\frac{1}{R_2}-\frac{2R_2}{r^2}\right)\right]\\
&+(\beta_2x_2^2-2\beta_1x_1x_2)D_5
\,,
\end{align*}
\begin{align*}
u^*_3(&\vc x)\mbox{}=-(\alpha_1x_3+2\beta_1 x_3)D_1+2L^2\alpha_1 x_3D_3
+\frac{\alpha_2x_2x_3D_2}{r^2}\\
&+\frac{\beta_2x_2x_3}{r^2}\left[(x_1-a)\left(\frac{1}{R_1}-\frac{2R_1}{r^2}\right)
-(x_1+a)\left(\frac{1}{R_2}-\frac{2R_2}{r^2}\right)\right]\\
&+\frac{2L^2\alpha_2x_2x_3}{r^2}\left[(x_1+a)\left(\frac{1}{R_1^3}+\frac{2}{R_1r^2}\right)
-(x_1-a)\left(\frac{1}{R_2^3}+\frac{2}{R_2r^2}\right)\right]\\
&+(\beta_2x_2x_3-2\beta_1x_1x_3)D_5\,,
\end{align*}
where $\alpha_1$, $\alpha_2$, $\beta_1$, and $\beta_2$ are constants to be determined.

We introduce prolate spheroidal coordinates $(u,v,\phi)$, defined by
\begin{gather*}
x_1=a\cosh u\cos v\,,\\
x_2=a\sinh u\sin v\cos\phi\,,\\
x_3=a\sinh u\sin v\sin\phi\,,
\end{gather*}
with $u\in[0,+\infty)$, $v\in[0,\pi]$, and $\phi\in[0,2\pi)$. 
The surfaces given by $u=c$, with $c>0$, are confocal prolate spheroids with foci at $x_1=\pm a$, and $u=0$ describes the degenerate spheroid $\rod$. 
Using these coordinates, we have
\begin{gather*}
r=a\sinh u\sin v\,,
\\
R_1=a(\cosh u+\cos v)\,,
\\
R_2=a(\cosh u-\cos v)\,,
\\
R_1-(x_1+a)=a(\cosh u-1)(1-\cos v)\,,
\\
R_2-(x_1-a)=a(\cosh u+1)(1-\cos v)\,,
\end{gather*}
and we can rewrite the components of $\vc u^*$ as
\begin{align*}
u^*_1(u,&v,\phi)=\mbox{}-2(\alpha_1+\beta_1)\log\left(\frac{\cosh u+1}{\cosh u-1}\right)\\
&-4\beta_1 \frac{\cosh u\cos^2 v}{\cosh^2 u-\cos^2 v}+2\alpha_1\frac{\cosh u\sin^2 v}{\cosh^2 u-\cos^2 v}\\
&+\frac{4L^2\alpha_1}{a^2}\frac{\cosh u\cos^2 v(3\cosh^2 u+\cos^2 v)}{(\cosh^2 u-\cos^2 v)^3}\\
&-2\alpha_2 \frac{\cos\phi\sinh u\sin v\cos v}{\cosh^2 u-\cos^2 v}\\
&+2\beta_2\frac{\cos\phi\cos v}{\sinh u\sin v}+\frac{4L^2\alpha_2}{a^2}\frac{\cos\phi\cos v}{\sinh u\sin v\left(\cosh^2 u-\cos^2 v\right)}\\
&+\frac{\cos\phi \cosh u\cos v(\cosh u\cos v+1)}{\sinh u\sin v}\left(\frac{2aL^2\alpha_2}{R_1^3}-\frac{a\beta_2}{R_2}\right)\\
&+\frac{\cos\phi \cosh u\cos v(\cosh u\cos v-1)}{\sinh u\sin v}\left(\frac{a\beta_2}{R_1}-\frac{2aL^2\alpha_2}{R_2^3}\right)\\
&+a(\beta_2 \sinh u \sin v\cos\phi-2\beta_1\cosh u\cos v)D_4\,,
\end{align*}
\begin{align*}
u^*_2(u,v,&\phi)=(\beta_2-\alpha_2)\log\left(\frac{\cosh u+1}{\cosh u-1}\right)\\
-&(2\alpha_1+4\beta_1)\frac{\cos\phi\sinh u\sin v\cos v}{\cosh^2 u-\cos^2 v}\\
+\mbox{}&\frac{4L^2\alpha_1}{a^2}\frac{\cos\phi\sinh u\sin v\cos v(3\cosh^2 u+\cos^2 v)}{(\cosh^2 u-\cos^2 v)^3}\\
+\mbox{}&\frac{\cosh u\cos v-1}{\sinh^2 u\sin^2 v}\left(\frac{\beta_2R_1}{a}+\frac{2L^2\alpha_2}{aR_2}\right)\\
-&\frac{\cosh u\cos v+1}{\sinh^2 u\sin^2 v}\left(\frac{\beta_2R_2}{a}+\frac{2L^2\alpha_2}{aR_1}\right)\\
+\mbox{}&2\alpha_2\frac{\cos^2 \phi\cosh u\sin^2 v}{\cosh^2 u-\cos^2 v}\left(1+\frac{4L^2}{a^2\sinh^2 u\sin^2 v}\right)\\
+&2L^2\alpha_2\cos^2 \phi\left(\frac{x_1+a}{R_1^3}-\frac{x_1-a}{R_2^3}\right)\\
+\mbox{}&\beta_2\cos^2 \phi\left[(x_1-a)\left(\frac{1}{R_1}-\frac{2R_1}{a^2\sinh^2 u\sin^2 v}\right)\right.\\
&\qquad\qquad\left.-(x_1+a)\left(\frac{1}{R_2}-\frac{2R_2}{a^2\sinh^2 u\sin^2 v}\right)\right]\\
+\mbox{}(\beta_2a^2&\cos^2 \phi\sinh^2 u\sin^2 v-2\beta_1 a^2\cos\phi\cosh u\cos v \sinh u\sin v)D_5
\,,
\end{align*}
\begin{align*}
u^*_3(u,v,\phi&)=-(2\alpha_1+4\beta_1)\frac{\sin\phi\sinh u\sin v\cos v}{\cosh^2 u-\cos^2 v}\\
+\mbox{}&\frac{4L^2\alpha_1}{a^2}\frac{\sin\phi\sinh u\sin v\cos v(3\cosh^2 u+\cos^2 v)}{(\cosh^2 u-\cos^2 v)^3}\\
+\mbox{}&+2\alpha_2\frac{\cos\phi\sin\phi\cosh u\sin^2 v}{\cosh^2 u-\cos^2 v}\\
+\mbox{}&\beta_2\cos\phi\sin\phi\left[(x_1-a)\left(\frac{1}{R_1}-\frac{2R_1}{a^2\sinh^2 u\sin^2 v}\right)\right.\\
&\qquad\qquad\qquad\left.
-(x_1+a)\left(\frac{1}{R_2}-\frac{2R_2}{a^2\sinh^2 u\sin^2 v}\right)\right]\\
+\mbox{}&2L^2\alpha_2\cos\phi\sin\phi(x_1+a)\left(\frac{1}{R_1^3}+\frac{2}{R_1a^2\sinh^2 u\sin^2 v}\right)\\
-\mbox{}&2L^2\alpha_2\cos\phi\sin\phi(x_1-a)\left(\frac{1}{R_2^3}+\frac{2}{R_2a^2\sinh^2 u\sin^2 v}\right)\\
+\mbox{}&D_5\beta_2a^2\cos\phi\sin\phi\sinh^2 u\sin^2 v\\
-\mbox{}&2D_5\beta_1 a^2\sin\phi\cosh u\cos v \sinh u\sin v\,.
\end{align*}
Notice that, in the limit $u\to 0$, $\vc u^*$ is divergent. Since we wish to impose the velocity of the fluid precisely on the set $\rod$ defined by $u=0$, we must add a suitable solution of the homogeneous equation associated with~\eqref{eq:helm}, thereby canceling the divergent terms.

We introduce external prolate spheroidal wave functions (using the notation of \cite{Erd55}) defined for $n\in\N$ and $l=0,\ldots,n$:
\[
W_n^{\pm l}(u,v,\phi)=S_n^{l(3)}(\cosh u,\kappa^2a^2/4)\mathrm{Ps}_n^l(\cos v,\kappa^2a^2/4)e^{\pm il\phi}\,.
\]
Those are solutions of $\lap W+\kappa^2 W=0$, regular at infinity but divergent on $\rod$. In our problem $\kappa=i/L$, and we are interested in the cases $n=0,1,2,3$. We have
\begin{align*}
W_0^{0}(u,&v,\phi)=S_0^{0(3)}(\cosh u,-a^2/4l^2)\mathrm{Ps}_0^0(\cos v,-a^2/4l^2)\\
=\mbox{}&s_0^0\sum_{2r\geq 0}\sum_{2p\geq 0}(-1)^p a_{0,r}^0a_{0,p}^0\sqrt{\frac{2L}{a\pi\cosh u}}\\
\times&e^{-i\pi r}K_{2r+\frac{1}{2}}\left(\frac{a\cosh u}{L}\right)P_{2p}^{0}(\cos v)
\,,
\end{align*}
\begin{align*}
W_1^{0}(u,&v,\phi)=S_1^{0(3)}(\cosh u,-a^2/4l^2)\mathrm{Ps}_1^0(\cos v,-a^2/4l^2)\\
=\mbox{}&s_1^0\sum_{2r\geq -1}\sum_{2p\geq -1}(-1)^p a_{1,r}^0a_{1,p}^0\sqrt{\frac{2L}{a\pi\cosh u}}\\
\times&e^{-i\pi(r+1/2)}K_{2r+\frac{3}{2}}\left(\frac{a\cosh u}{L}\right)P_{2p+1}^{0}(\cos v)
\,,
\end{align*}
\begin{align*}
W_1^{\pm 1}(u,&v,\phi)=S_1^{1(3)}(\cosh u,-a^2/4l^2)\mathrm{Ps}_1^1(\cos v,-a^2/4l^2)e^{\pm i\phi}\\
=\mbox{}&e^{\pm i\phi}\left(1-\frac{1}{\cosh^{2} u}\right)^{\frac{1}{2}}s_1^{-1}\times\\
\times&\sum_{2r\geq 0}\sum_{2p\geq 0}(-1)^p a_{1,r}^{-1}a_{1,p}^1\sqrt{\frac{2L}{a\pi\cosh u}}\\
\times&e^{-i\pi(r+1/2)}K_{2r+\frac{3}{2}}\left(\frac{a\cosh u}{L}\right)P_{2p+1}^{1}(\cos v)
\,,
\end{align*}
\begin{align*}
W_2^{0}(u,&v,\phi)=S_2^{0(3)}(\cosh u,-a^2/4l^2)\mathrm{Ps}_2^0(\cos v,-a^2/4l^2)\\
=\mbox{}&s_2^0\sum_{2r\geq -2}\sum_{2p\geq -2}(-1)^p a_{2,r}^0a_{2,p}^0\sqrt{\frac{2L}{a\pi\cosh u}}\\
\times&e^{-i\pi(r+1)}K_{2r+\frac{5}{2}}\left(\frac{a\cosh u}{L}
\right)P_{2p+2}^{0}(\cos v)
\,,
\end{align*}
\begin{align*}
W_2^{\pm 1}(u,&v,\phi)=S_2^{1(3)}(\cosh u,-a^2/4l^2)\mathrm{Ps}_2^1(\cos v,-a^2/4l^2)e^{\pm i\phi}\\
=\mbox{}&e^{\pm i\phi}\left(1-\frac{1}{\cosh^{2} u}\right)^{\frac{1}{2}}s_2^{-1}\times\\
\times&\sum_{2r\geq -1}\sum_{2p\geq -1}(-1)^p a_{2,r}^{-1}a_{2,p}^1\sqrt{\frac{2L}{a\pi\cosh u}}\\
\times&e^{-i\pi(r+1)}K_{2r+\frac{5}{2}}\left(\frac{a\cosh u}{L}\right)P_{2p+2}^{1}(\cos v)
\,,
\end{align*}
\begin{align*}
W_2^{\pm 2}(u,&v,\phi)=S_2^{2(3)}(\cosh u,-a^2/4l^2)\mathrm{Ps}_2^2(\cos v,-a^2/4l^2)e^{\pm i2\phi}\\
=\mbox{}&e^{\pm i2\phi}\left(1-\frac{1}{\cosh^{2} u}\right)s_2^{-2}\times\\
\times&\sum_{2r\geq 0}\sum_{2p\geq 0}(-1)^p a_{2,r}^{-2}a_{2,p}^2\sqrt{\frac{2L}{a\pi\cosh u}}\\
\times&e^{-i\pi(r+1)}K_{2r+\frac{5}{2}}\left(\frac{a\cosh u}{L}\right)P_{2p+2}^{2}(\cos v)
\,,
\end{align*}
\begin{align*}
W_3^{0}(u,&v,\phi)=S_3^{0(3)}(\cosh u,-a^2/4l^2)\mathrm{Ps}_3^0(\cos v,-a^2/4l^2)\\
=\mbox{}&s_3^0\sum_{2r\geq -3}\sum_{2p\geq -3}(-1)^p a_{3,r}^0a_{3,p}^0\sqrt{\frac{2L}{a\pi\cosh u}}\\
\times&e^{-i\pi(r+3/2)}K_{2r+\frac{7}{2}}\left(\frac{a\cosh u}{L}\right)P_{2p+3}^{0}(\cos v)
\,,
\end{align*}
\begin{align*}
W_3^{\pm 1}(u,&v,\phi)=S_3^{1(3)}(\cosh u,-a^2/4l^2)\mathrm{Ps}_3^1(\cos v,-a^2/4l^2)e^{\pm i\phi}\\
=\mbox{}&e^{\pm i\phi}\left(1-\frac{1}{\cosh^{2} u}\right)^{\frac{1}{2}}s_3^{-1}\times\\
\times&\sum_{2r\geq -2}\sum_{2p\geq -2}(-1)^p a_{3,r}^{-1}a_{3,p}^1\sqrt{\frac{2L}{a\pi\cosh u}}\\
\times&e^{-i\pi(r+3/2)}K_{2r+\frac{7}{2}}\left(\frac{a\cosh u}{L}\right)P_{2p+3}^{1}(\cos v)
\,,
\end{align*}
\begin{align*}
W_3^{\pm 2}(u,&v,\phi)=S_3^{2(3)}(\cosh u,-a^2/4l^2)\mathrm{Ps}_3^2(\cos v,-a^2/4l^2)e^{\pm i2\phi}\\
=\mbox{}&e^{\pm i2\phi}\left(1-\frac{1}{\cosh^{2} u}\right)s_3^{-2}\times\\
\times&\sum_{2r\geq -1}\sum_{2p\geq -1}(-1)^p a_{3,r}^{-2}a_{3,p}^2\sqrt{\frac{2L}{a\pi\cosh u}}\\
\times&e^{-i\pi(r+3/2)}K_{2r+\frac{7}{2}}\left(\frac{a\cosh u}{L}\right)P_{2p+3}^{2}(\cos v)
\,,
\end{align*}
where $s_n^{l}$ and $a_{n,m}^{l}$ are coefficients dependent on $\kappa^2 a^2/4=-a^2/4l^2$, $K_{\nu}$ is a modified Bessel function of the second kind, and $P_{\nu}^{\mu}$ is an associated Legendre function of the first kind.

The divergence of the logarithmic terms in $u^*_1$ and $u^*_2$ can be cancelled by adding multiples of $W_0^0$. If $U_2=\alpha_2=\beta_2=0$, which corresponds to a uniform translation of the rod along its axis, the dependence on $\phi$ and the radial decay of $\vc u^*$ suggest that a solution $\vc u^{\parallel}$ of~\eqref{eq:helm} can be written in the form
\begin{align*}
u^{\parallel}_1&\mbox{}=u^*_1+c_1W_0^0+d_1W_1^0+k_1W_2^0+e_1W_3^0  \,,\\
u^{\parallel}_2&\mbox{}=u^*_2+c_2(W_1^1+W_1^{-1})+d_2(W_2^1+W_2^{-1})+k_2(W_3^1+W_3^{-1})   \,,\\
u^{\parallel}_3&\mbox{}=u^*_3+ic_3(W_1^1-W_1^{-1})+id_3(W_2^1-W_2^{-1})+ik_3(W_3^1-W_3^{-1})  \,.
\end{align*}
The no-slip condition on $\rod$ requires that $\vc u(0,v,\phi)=-\vc U$, which fixes the value of $\alpha_1$, once the boundedness of $\vc u^{\parallel}$ is ensured by a suitable choice of the the coefficients $c_i$, $d_i$, $k_i$, and $e_1$ as functions of $\alpha_1$.

On the other hand, in the case $U_1=\alpha_1=\beta_1=0$, corresponding to a translation of the rod with direction in the plane orthogonal to $\vc e_1$, we may seek a solution $\vc u^{\perp}$ of~\eqref{eq:helm} of the form
\begin{align*}
u^{\perp}_1&\mbox{}=u^*_1+c_1(W_1^1+W_1^{-1})+d_1(W_2^1+W_2^{-1})+k_1(W_3^1+W_3^{-1})   \,,\\
u^{\perp}_2&\mbox{}=u^*_2+c_2W_0^0+d_2(W_2^2+W_2^{-2}+A_2W_2^0+B_2W_0^0)\\
&\quad+k_2(W_3^2+W_3^{-2}+A_3W_3^0+B_3W_0^0) \,,\\
u^{\perp}_3&\mbox{}=u^*_3+id_3(W_2^2-W_2^{-2})+ik_3(W_3^2-W_3^{-2}) \,.
\end{align*}
As above, the no-slip condition on $\rod$ fixes the value of $\alpha_2$, once the boundedness of $\vc u^{\perp}$ is ensured by a suitable choice of the coefficients $A_i$, $B_i$, $c_i$, $d_i$, and $k_i$. Finally, the disturbance field $\vc u$ for a uniform translation with generic velocity $\vc U=U_1\vc e_1+U_2\vc e_2$ is given by $\vc u=\vc u^{\parallel}+\vc u^{\perp}$.

Notice that, since we can impose the corresponding boundary conditions on the surface described by $u=c$, for any value $c \geq  0$, the previous expressions provide the general solution for the flow past any translating spheroid of the family with foci at $x_1=\pm a$.


\end{document}